\documentclass[aps,pre,twocolumn]{revtex4}
\usepackage{graphicx,xcolor}
\usepackage{subcaption}
\usepackage{pdfpages}
\usepackage{makecell}
\newcommand\xoutpars[1]{\let\helpcmd\xout\parhelp#1\par\relax\relax}
\newcommand\soutpars[1]{\let\helpcmd\sout\parhelp#1\par\relax\relax}
\long\def\parhelp#1\par#2\relax{%
  \helpcmd{#1}\ifx\relax#2\else\par\parhelp#2\relax\fi%
}
\usepackage{amssymb,amsfonts,amsmath,bm,hyperref}
\usepackage[latin1]{inputenc}
\usepackage{makeidx}
\usepackage{graphicx}
\hypersetup{
	colorlinks,
	linkcolor={magenta},
	citecolor={purple},
	urlcolor={violet}
}

\usepackage[T1]{fontenc}
\usepackage{ulem} 
\begin{document}
    \setlength{\parskip}{0pt}
\title{\bf{Modelling cargo transport in  crowded environments: effect of motor 
association  to cargos}}

\author{Sutapa Mukherji${}$}
\email{ sutapa.mukherji@ahduni.edu.in}
\author{Dhruvi K. Patel}
\affiliation{Mathematical and Physical Sciences division, School of Arts and Sciences, Ahmedabad University, Navrangpura, 
Ahmedabad 380009, India}
\date{\today}
\begin{abstract}
In intracellular transports, motor proteins transport   macromolecules as cargos 
to  desired  locations by moving on 
 biopolymers such as microtubules. 
 Recent experiments  suggest that cargos that  
can  associate motor proteins during their translocation have larger
 run-length, association time and  can overcome the motor traffic on   microtubule tracks.  
Here, we model the dynamics of a cargo that can associate at the most $m$ 
free motors present on the track  as obstacles to its  motion. 
 The proposed models display
  competing effects of association  and  crowding, leading to a peak in 
  the run-length with  the free motor density. This result is consistent  with 
  past experimental observations.
For $m=2$ and $3$, we show that this feature is governed by the largest eigenvalue of 
the  transition matrix describing the cargo dynamics. 
 In all the above cases, free motors are assumed 
to be present as stalled obstacles.  We finally compare  simulation results for the 
run-length for   general scenarios where the free motors undergo processive motion in addition 
to binding and unbinding to or from  the microtubule.   
\end{abstract}
\maketitle
\section{Introduction}
Intracellular transport  often involves   directional  movements of motor proteins 
 on biopolymers
such as microtubules or  actin filaments  \cite{howard1,schliwa}. Three major classes of 
motor proteins known as kinesin, dynein and myosin are responsible for such   transports.   
Using the energy derived from  the  hydrolysis of adenosinetriphosphate (ATP)  molecules, motor 
proteins   transport different types of 
cargos such as  cellular organelles,  protein complexes, mRNAs etc.  to 
desired locations in the cell. Such cargo movements are essential for various cellular functions 
 such as cell morphogenesis, cell division, cell growth etc. This motion is processive in the 
sense that motor proteins typically  move over several  successive 
  steps before detaching from the microtubule.
  Early studies  \cite{block,cross,howard2} on intracellular 
 transport  revealed the underlying mechanism behind motor transport and 
how various properties such as the run-length, velocity etc. depend, for example, 
on  the external force or the 
 concentration of ATP molecules.
While many of these studies are around  the transport by a single motor, it is believed that cargos 
are often transported by  multiple motors \cite{gross,holzwarth,gelfand,unger} 
which help cargos remain bound to the biopolymer for 
a longer time.  Experimental and theoretical studies \cite{holzwarth,klumpp,beeg} show that 
the presence  of several motors helps the cargo overcome the viscous drag of the cytoplasm and have 
larger velocity as compared to transport by single motors. The cooperation of several motors also  
leads   to  longer run-length  of the cargo before it  detaches  from the microtubule. Further, in vitro 
experiments indicate that  transport processes by multiple motors can be efficiently regulated by controlling the 
number of engaging motors \cite{vershinin}. Besides these studies, there have been extensive
 experimental and theoretical studies attempting to understand the collective nature of  transports 
involving many motors under diverse conditions 
\cite{leduc,segregation,clogging, helical,tasep1,multi-opposite,frey-lang,multi-same,laneswitch,gov,freygait}.
 
   
 Quite often such transport processes take place in a crowded environment of  
 the intracellular space. This is in particular true for the  axon region of the neuron  
 cell where a dense network of biopolymers,  pre-exisiting organelles and the narrow geometry of 
 the axon  together give rise to a crowded environment that can impede cargo movements. 
 However, despite  crowding, it is found that the cargo transport happens in a robust manner without 
 significant jamming or cargo dissociation. Experiments elucidating cargo transport in crowded 
 environments indicate that motor proteins can  adapt alternative strategies that 
 might  help  them circumvent the crowding problem \cite{leduc,surrey}. 
  In a recent experimental study  aimed at understanding  the motion of  a cargo 
 in a crowded environment, Conway {\it et al} \cite{conway,conway1} studied 
  the motile properties  of  quantum dot (Qdot) cargos, that can associate multiple kinesins, 
 on a microtubule crowded with free kinesin motors.   While comparing the motile properties of free kinesins 
 and the  Qdot cargos in crowded conditions, cargos were found to display 
  longer run-lengths and association times as 
   compared to free kinesins as the motor density increased. 
   This  difference prompted the prediction that 
  the property of a cargo to associate multiple motors   helps increase its run-length,  
  association time and overcome the motor traffic. 
 It was observed that while translocating, Qdot cargos could associate kinesins from 
   the microtubule pool, dissociate kinesins attached to itself, or associate kinesins that are already 
   moving along the microtubule and move together subsequently.

 Motivated by this work,  here, we  propose 
 mathematical and computational models  to characterise the motion 
 of a single cargo on a track crowded by free kinesin motors. 
 During its translocation, the cargo 
 can associate  free motors which    impede the motion of the cargo by occupying 
the  forward sites on the microtubule.  
We assume that upon such association, a  kinesin detaches itself from the microtubule 
rendering  the forward site free. Our aim is to find how the interplay of the 
kinesin-association property of the  cargo and the crowding along the track affects 
cargo's motile properties, for example, its run-length and association time etc.  
To  our knowledge, this is  the first modelling study of cargo transport where 
the cargo has the ability to associate  kinesins present on  a  crowded 
microtubule track.

\begin{figure}[!]
	\includegraphics[width=0.45\textwidth]{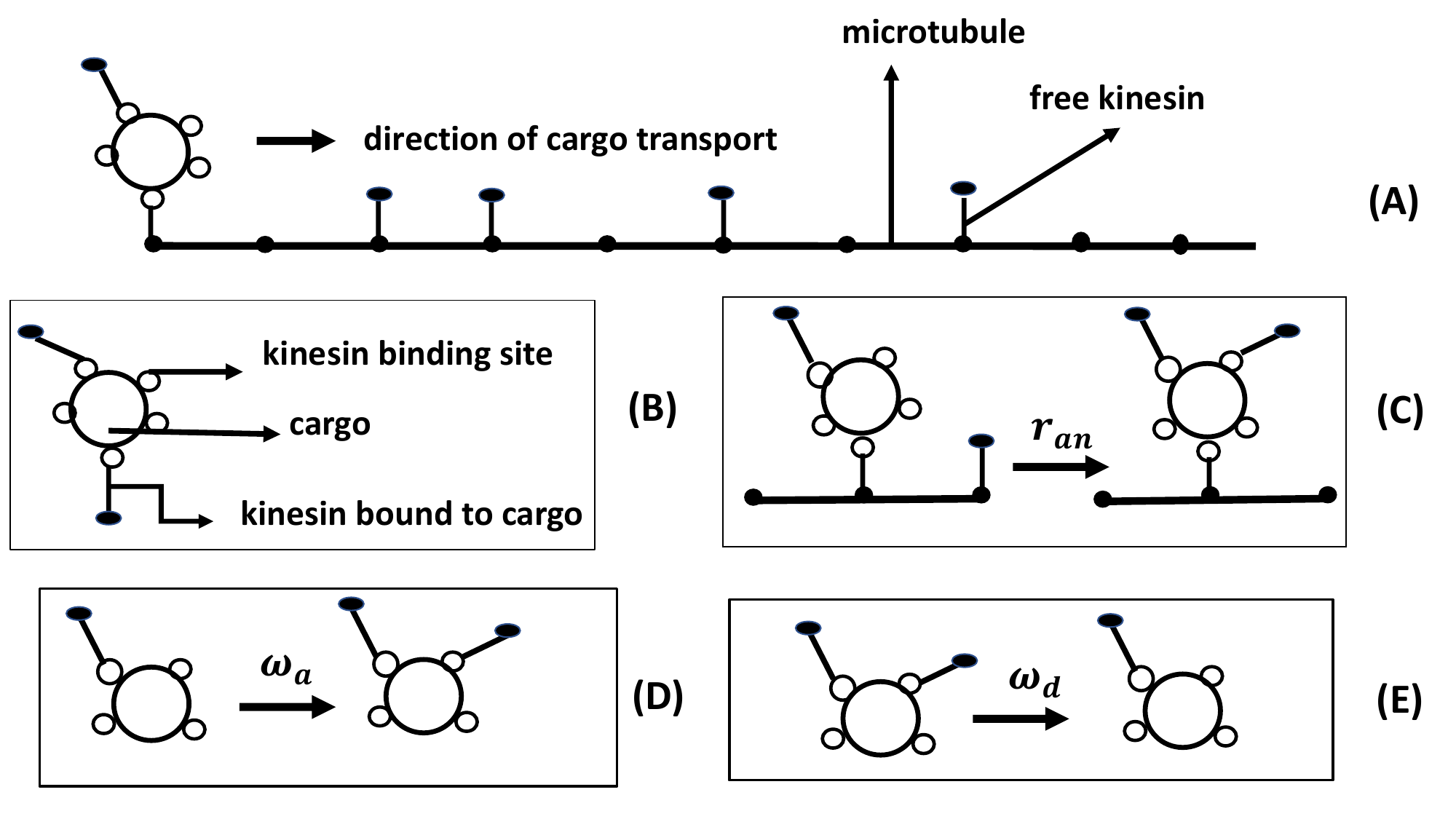}
	\caption{(A)  Cargo transport on a microtubule in the presence of free kinesins bound to microtubules. 
	The microtubule is  represented by a one-dimensional lattice with a site representing a tubulin dimer, 
	the basic subunit of a  microtubule. 
	(B)  A cartoon of a cargo bound to two kinesins. The cargo has 
	more than one kinesin binding sites. (C) The process of kinesin association to cargo at  
	rate $r_{an}$. (D) The process of attachment of a kinesin from the intracellular space to the cargo at 
	rate $\omega_a$.
	(E) The process of kinesin detachment from the cargo at rate $\omega_d$.}
	\label{fig:modelfig}
\end{figure}

To this end, we study cargo transport under different scenarios described below. 
 (1) In the simplest scenario, we assume that 
 the cargo is  always bound to the microtubule. 
 Along the path of the cargo, 
 the microtubule binding sites are randomly occupied by free kinesin motors. 
  We assume that the  kinesin that gets associated to the cargo during its translocation
 plays no specific role in facilitating the forward motion of the cargo 
 other than freeing the forward site. 
 This is equivalent to assuming  that the cargo removes 
 the kinesin molecule occupying the forward site via the association process. 
 For this model, referred as "Model 1" below, 
 we find the average velocity of the cargo.   
  (2) In model 2, we assume that the cargo can  bind 
 more than one, say, at the most $m$ number of  kinesins. This is based on the 
 predictions  that the cargo may have a finite number of kinesin binding sites \cite{conway}.  
 Hence, the cargo can associate a kinesin occupying the forward site 
 provided  it has a free binding site available.  
 We  consider  $m=2,\ 3, {\rm and} \ 4$ in the following analysis.    Kinesins 
 attached to the cargo can detach from it  and a free kinesin from the intracellular 
 space can attach  to the cargo at given  rates.   Finally, we implement the  condition 
  that a cargo can no longer  be 
 on the microtubule track if all the kinesin molecules detach from the cargo. 
A generalised version of the  mathematical formulation of  model 1
 allows us 
to analyse the  cargo motion obeying above rules for  $m=2$ and $3$.  
 Finally, run-lengths of  $m=2, \ 3, {\rm and\ 4}$ are found 
 upon numerically simulating the cargo dynamics. 
  The motion of the 
 cargo following different dynamical rules are shown in figure \ref{fig:modelfig}.
(3) In models 1 and 2,  free kinesins are assumed  to be stalled on the microtubule. 
In model 3, 
we simulate  cargo dynamics in the presence of moving kinesins as well as random processes of 
kinesin binding and unbinding to or from the microtubule. 
We compare the  run-length of the cargo (with $m=3$)  in the presence or absence of 
various processes mentioned above.

\section{Models and Results}
\subsection{Model 1}
The motion of the cargo is modelled considering the following dynamical rules. 
(a) The cargo transported by  a kinesin starts its forward  
journey from a given point on a one-dimensional track (often referred below as a lattice) 
representing the microtubule. 
(b) For all the lattice sites ahead,  we assume  an initial, random distribution of free 
kinesins. The  average kinesin density  on the lattice is represented  by $r_m$. These kinesins are 
assumed to be stalled. 
(c) The cargo moves to the forward site provided the forward site is not occupied 
by a free kinesin. 
(d) If the forward site in front of the cargo  is  occupied by a  free 
kinesin, the cargo  can associate  the kinesin  with itself at  
rate $r_{an}$ rendering the forward site free.  

\begin{figure}[!]
	\includegraphics[width=0.45\textwidth]{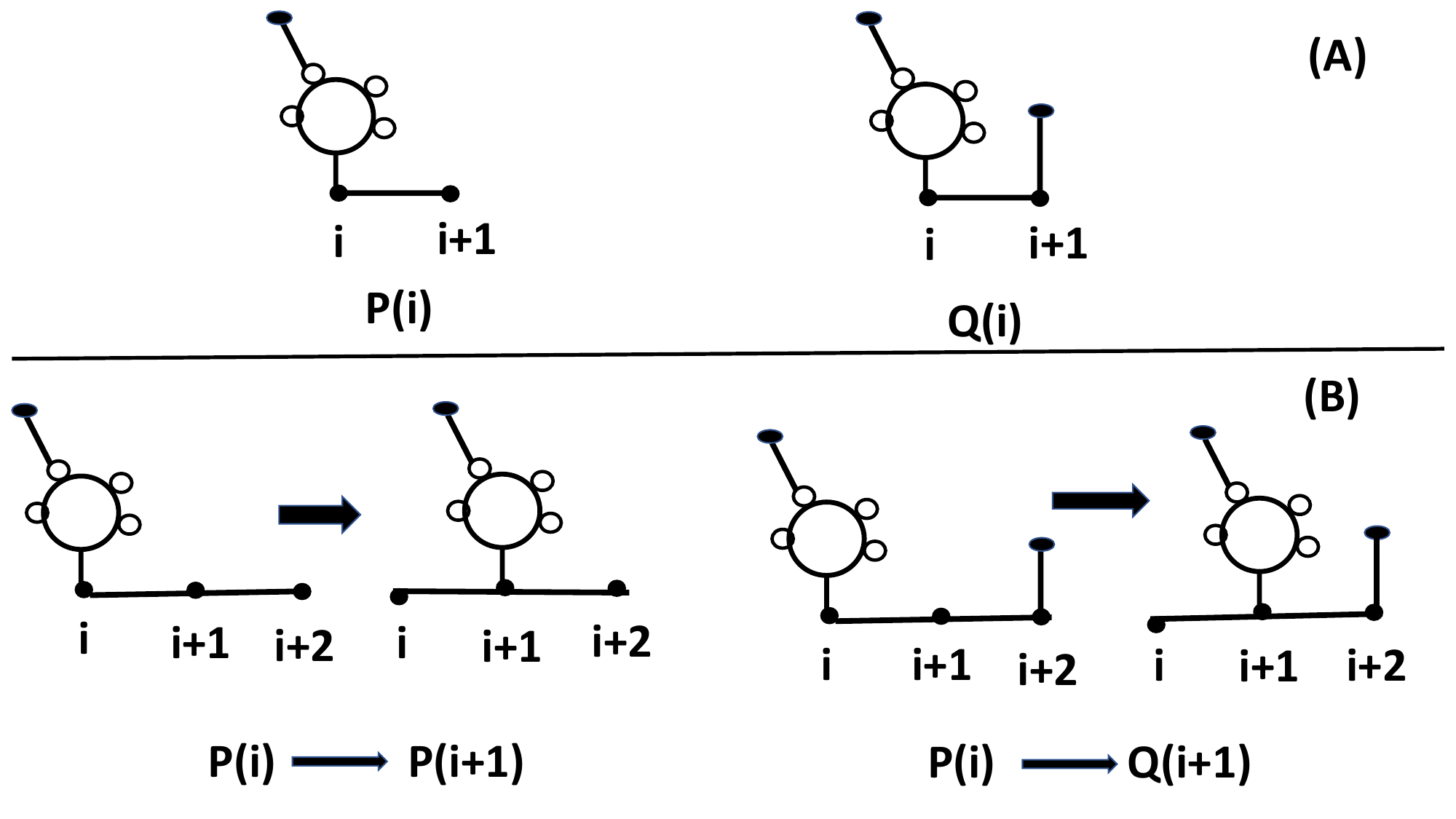}
	\caption{(A) Two possible configurations of two neighbouring sites with the first site  occupied by 
	the motor transporting the cargo. (B) Possible transitions associated with the transport of the 
	cargo to the next site.}
	\label{fig:config}
\end{figure}

In order to  build the mathematical model,  we consider possible configurations that  two 
neighbouring sites can have  when the first one of them is  occupied by the cargo. 
 For $i$  and $(i+1)$-th sites,  with the  cargo being at the $i$-th site, the   $(i+1)$-th site can be   
either empty or occupied by a free kinesin molecule. We denote the 
 probability of finding $(i+1)$-th site empty  with the 
$i$-th site occupied by the cargo at time $t$ by $P(i,t)$. Similarly,  
the probability of finding   $(i+1)$-th site 
occupied by a free kinesin while  the cargo  is at the $i$-th site at time $t$  is $Q(i,t)$. 
 Figure (\ref{fig:config}) shows these configurations as well as possible transitions from one 
 configuration  to the other as the cargo translocates forward.
The following equations describe how these two configurations evolve with time \cite{sm}. 
\begin{eqnarray}
&&\frac{dP(i)}{dt}=r_{an} Q(i)+(1-r_m) P(i-1)-
P(i), \nonumber\\ \label{pevolve}\\
&&\frac{dQ(i)}{dt}=-r_{an} Q(i)+r_m P(i-1). \label{qevolve}
\end{eqnarray}
 The first term  on the RHS of equation  (\ref{pevolve})
indicates a cargo-association process due to which a Q-type 
configuration transitions to a $P$-type configuration.  The term with the pre-factor 
 $(1-r_m)$ indicates the motion of the cargo from  
  $(i-1)$-th site to  
$i$-th site while the $(i+1)$-th site is vacant. While the forward motion happens with unit rate, 
the factor $(1-r_m)$ indicates the probability that after 
the forward motion, $(i-1)\rightarrow i$, 
the cargo lands in a P-type configuration i.e. the $(i+1)$-th site is 
unoccupied by a kinesin.   The last term in (\ref{pevolve}) is a loss term which 
indicates  that a cargo has moved from the $i$-th site to the $(i+1)$-th site. 
In equation (\ref{qevolve}), the first term on the RHS indicates a 
loss of a $Q$-type configuration due to the association process. The second term is a gain term due to 
the hopping of the cargo from $(i-1)\rightarrow i$  where $(i+1)$-th site is occupied by a free kinesin. 

To  find the average properties of the cargo motion,  
we define   generating functions corresponding to the two probabilities  as  
\begin{eqnarray}
\tilde P(\gamma)=\sum_{i=-\infty}^{\infty} \gamma^i P(i), \ {\rm and} \  
\tilde Q(\gamma)=\sum_{i=-\infty}^{\infty}\gamma^i Q(i).
\end{eqnarray}
In terms of these generating functions, the time evolution equations are 
\begin{eqnarray}
&&\frac{d}{dt}\tilde P(\gamma)=r_{an} \tilde Q(\gamma)+
(1-r_m) \gamma \tilde P(\gamma) - \tilde P(\gamma), \ {\rm and} \nonumber\\ \label{Peq}\\
&&\frac{d}{dt}\tilde Q(\gamma)=-r_{an} \tilde Q(\gamma)+r_m \gamma \tilde P(\gamma).\label{Qeq}
\end{eqnarray}
The average position of the cargo  can be found from the probabilities as 
\begin{eqnarray}
\langle i\rangle=\sum_{i=-\infty}^{\infty} i [P(i)+Q(i)] =\frac{d}{d\gamma}[\tilde P(\gamma)+\tilde Q(\gamma)]\mid_{\gamma=1}.
\end{eqnarray}
The average velocity of the cargo    is obtained from 
$v=\langle i\rangle/t$ where $t$ is the time taken to travel  an 
average distance $\langle i\rangle$. 
Solving equations (\ref{Peq}) and (\ref{Qeq}), the   average velocity of the 
cargo is found as (see Appendix \ref{appendix:model1} for details) 
\begin{eqnarray}
 v=\frac{r_{an}}{r_{an}+r_m}. \label{v-one-kinesin}
 \end{eqnarray}

\subsection{Model 2}
Here we generalize the mathematical framework, discussed in the previous section,  for higher 
values of $m$ taking into account the possibilities of  detachments of the cargo from the microtubule. In the following, we discuss the mathematical model   for  $m=2$ and simulation results
for $m=2, \ 3, \ {\rm and} \ 4$. The cargo dynamics for $m=3$ is discussed  in 
Appendix \ref{appendix:model2}. 

For $m=2$, the cargo  has two kinesin binding sites. 
Hence  it  can associate at the most two  kinesins.  
The basic  rules for cargo transport in this case are listed below.
(a) As before, we begin with an initial, random distribution of  stalled free kinesins 
on a one-dimensional lattice. The average density of free kinesins is $r_m$. 
(b) The cargo attached to a kinesin starts its forward 
journey from a given point on the lattice. 
(c) If the forward site is blocked by a free kinesin, the cargo  can associate the  
kinesin with itself at a rate   $r_{an}$ provided the cargo has only one kinesin 
bound  to it. 
(d) A kinesin bound to the  cargo can detach from the cargo at  rate $\omega_d$ 
and  a free kinesin from the intracellular space  can bind to the cargo at  
rate $\omega_a$ provided the cargo has only one kinesin attached to it. 
(e) A cargo is not attached to the microtubule if all the kinesins detach from the cargo. 
Thus, we are not being specific about how many kinesins are actively 
transporting the cargo or how many remain bound to the cargo without participating in 
 cargo transport  actively.

As before,  we begin with  two possible configurations of, say, $\{i, i+1\}$-th sites 
where $i$-th site is occupied by the cargo.  However, 
here the cargo can be  in two possible states  -  bound to one kinesin  or bound to two kinesins. 
 Hence, the  probabilities are defined in  the following way. $P_n(i,t)$ ($n=1,\ 2)$ represents the 
 probability, at time $t$, 
 of  the configuration where the cargo, located at $i$-th site,  is bound to $n$ kinesins  
 and the $(i+1)$-th site is empty.  
 Similarly, $Q_n(i,t)$ ($n=1,\ 2$)
  represents the probability, at time $t$,  of the configuration where the cargo, located 
 at $i$-th site, is  bound to $n$ kinesins and the $(i+1)$-th site is occupied 
  by a free kinesin. 
    
  The probabilities of various  configurations change with time as per the following equations,
\begin{widetext}
\begin{eqnarray}
&&\frac{d}{dt}P_2(i)=(1-r_m) P_2(i-1)-P_2(i)+r_{an}Q_1(i)-\omega_d P_2(i)+\omega_aP_1(i),\label{p2}\\
&&\frac{d}{dt}P_1(i)=(1-r_m) P_1(i-1)-
P_1(i)-\omega_a P_1(i)+\omega_d (P_2(i)- P_1(i)),\label{p1}\\
&&\frac{d}{dt}Q_2(i)=r_m P_2(i-1)+\omega_a Q_1(i)-\omega_d Q_2(i), \ {\rm and}\label{q2}\\
&&\frac{d}{dt}Q_1(i)=r_m P_1(i-1)-r_{an} Q_1(i)+\omega_d (Q_2(i)-Q_1(i))-
\omega_a Q_1(i).\label{q1}
\end{eqnarray}
\end{widetext}
The $r_{\rm an}$-dependent term in equation (\ref{p2}) represents a process of kinesin 
association by the cargo. Due to this process, a $Q_1$-type configuration transitions to a
$P_2$-type configuration.
$\omega_a$($\omega_d$) dependent terms represent kinesin attachment(detachment) processes 
to(from) the cargo. For example, the $\omega_d$ dependent term in equation (\ref{p2}) represents 
 detachment of a kinesin due to which the cargo transitions from  $P_2$ state to $P_1$ state.
In addition to above equations, we introduce  
probabilities $P_0(i,t)$ and $Q_0(i,t)$ of having  situations where 
  the cargo bound to one kinesin residing at $i$-th site at time $t$ loses its kinesin. 
  These  probabilities change with time  as per the  equations
\begin{eqnarray}
\frac{d}{dt} P_0(i)=\omega_d P_1(i)\label{p0} \ {\rm and}\ 
\frac{d}{dt} Q_0(i)=\omega_d Q_1(i) \label{q0}.
\end{eqnarray} 
Defining generating functions as $\tilde P_n(\gamma,t)=
\sum_{i=-\infty}^{\infty}\gamma^i P_n(i,t)$,  and 
$\tilde Q_n(\gamma,t)=\sum_{i=-\infty}^{\infty}\gamma^i Q_n(i,t)$ (where $n=0,\ 1,\ 2$), 
we can rewrite  equations (\ref{p2})-(\ref{q1})  as 
\begin{eqnarray}
\frac{d}{dt} {\bf H}(\gamma,t)={\bf S} {\bf H}(\gamma,t), 
\end{eqnarray}
where ${\bf H}$ is a column matrix 
\begin{eqnarray}
{\bf H}(\gamma,t)={\begin{pmatrix}
 & \tilde P_2(\gamma,t)\\
 &\tilde P_1(\gamma,t)\\
 & \tilde Q_2(\gamma,t)\\
 &\tilde Q_1(\gamma,t)\\
 \end{pmatrix} } 
  \end{eqnarray}
  and ${\bf S}$ is a $4\times 4$ matrix 
  \begin{widetext}
  \begin{eqnarray}
  {\bf S}={\begin{pmatrix}
   & (1-r_m) \gamma-1-\omega_d & \omega_a \  & 0  & r_{an} \\
 &\omega_d & (1-r_m) \gamma-1-\omega_a - \omega_d & 0 & 0\\
  & r_m\gamma & 0  & -\omega_d & \omega_a\\
 &0  & r_m\gamma & \omega_d   & -(\omega_a+\omega_d+r_{an}) \\
  \end{pmatrix} }.
 \end{eqnarray}
 \end{widetext}

\begin{figure}[!]
	\includegraphics[width=0.45\textwidth]{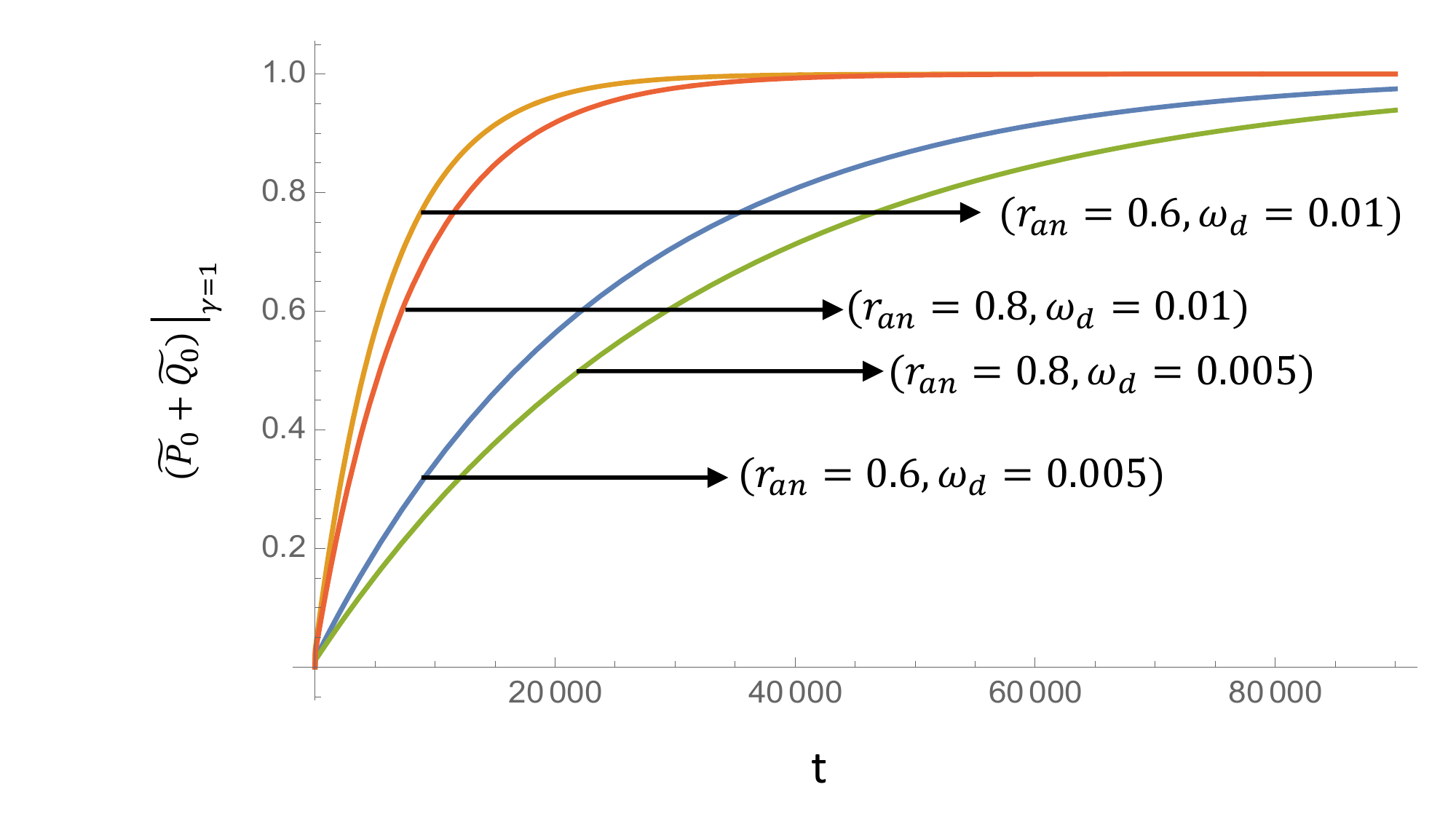}
	\caption{$y-$axis represents the total probability 
	($\sum_{i=-\infty}^\infty[P_0(i,t) +Q_0(i,t)]$) of the cargo losing its last kinesin 
	at time $t$ while being at any site on the lattice. For this plot, 
	$r_m=0.5$ and $\omega_a=0.01$.}
	\label{fig:vary-omega-d}
\end{figure}

\begin{figure}[!]
	\includegraphics[width=0.45\textwidth]{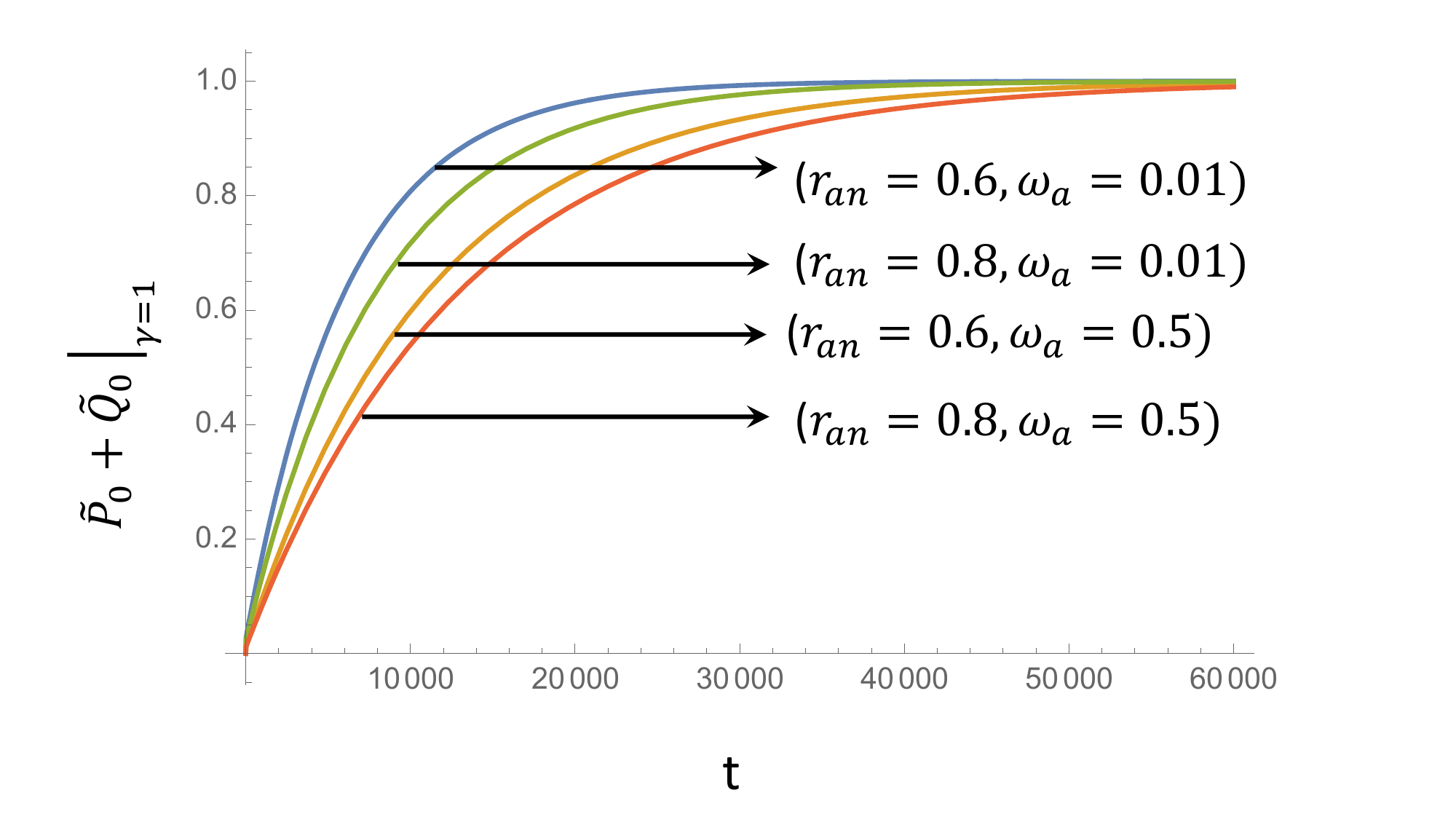}
	\caption{$y-$axis represents the total probability 
	($\sum_{i=-\infty}^\infty[P_0(i,t) +Q_0(i,t)]$) of the cargo losing its last kinesin 
	at time $t$ while being at any site on the lattice. For this plot, $r_m=0.5$ and $\omega_d=0.01$.}
	\label{fig:vary-omega-a}
\end{figure}

 
In order to have an estimate of  the association  time  of the cargo 
and how it is impacted by various processes, we have studied 
the quantity $[\tilde P_0(\gamma,t)+\tilde Q_0(\gamma,t)]\mid_{\gamma=1}$. 
This quantity being identical to $\sum_{i=-\infty}^\infty[P_0(i,t) +Q_0(i,t)]$ 
indicates the total probability of cargo being left with no kinesin bound 
 to it while being at any point on the lattice. A plot of this quantity for different 
 parameter  values  are shown in figures (\ref{fig:vary-omega-d}) and (\ref{fig:vary-omega-a}). 
 Over large time, this quantity approaches unity indicating cargo losing all the kinesins leading to the  
 detachment of  the cargo from the microtubule. The approach of this quantity to unity 
 is what provides us with an estimate of the association time of the cargo to the microtubule. 
 A fast approach to unity 
 indicates a low association time of the cargo. For both figures, 
we have chosen  the same sets of values for the kinesin-association rate, $r_{an}$. 
The increase or decrease in 
$\omega_a$ and $\omega_d$, respectively,  is expected to increase the association time of the cargo. 
Figures show  that  reducing  the kinesin detachment rate, $\omega_d$, from the cargo has  much 
stronger effects on the association  time  as compared to increasing the kinesin 
attachment rate, $\omega_a$. 

\begin{figure}[!]
	\includegraphics[width=0.55\textwidth]{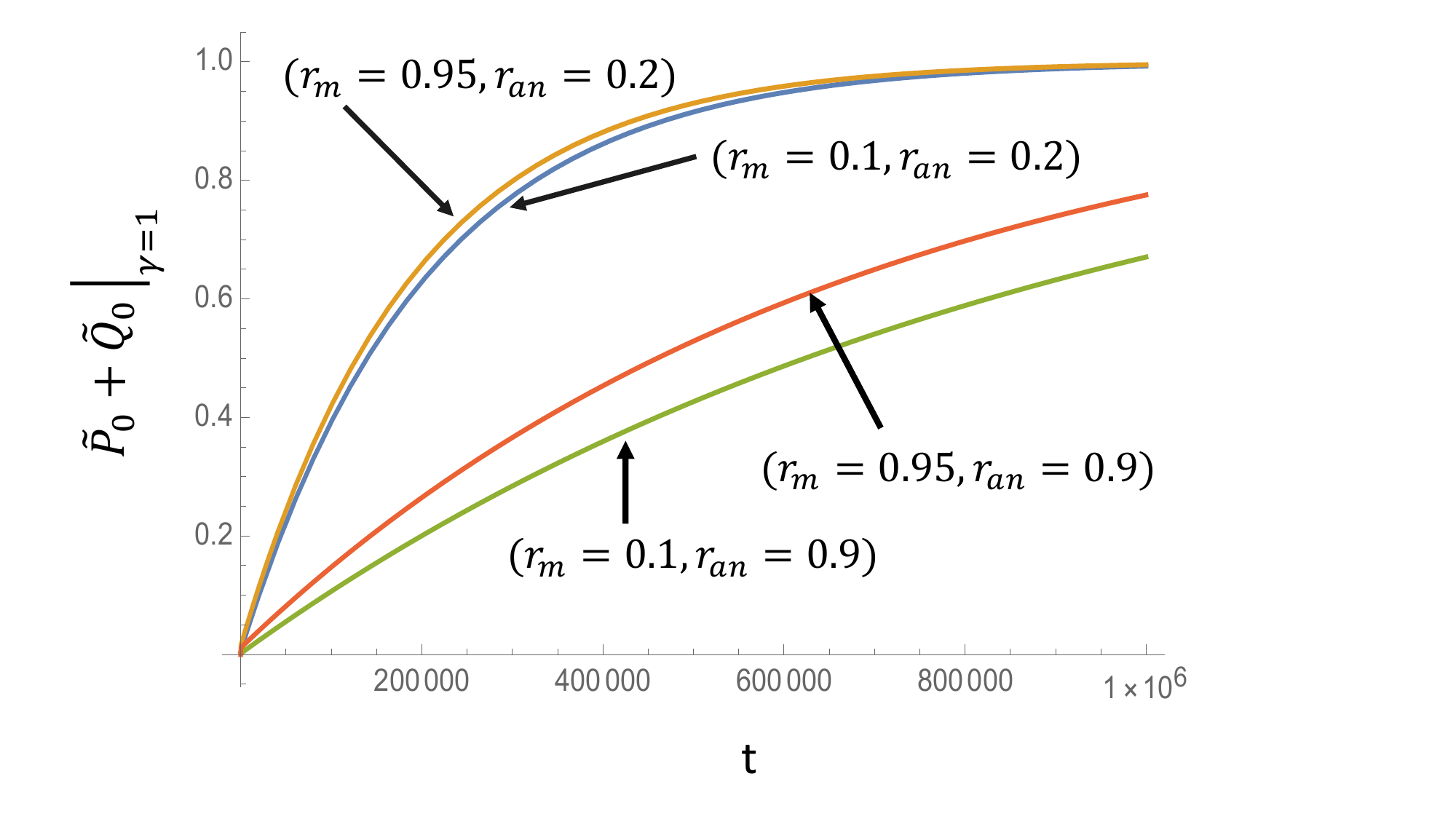}
	\caption{ $y-$axis represents the total probability 
	($\sum_{i=-\infty}^\infty[P_0(i,t) +Q_0(i,t)]$) of the cargo losing its last kinesin 
	at time $t$ while being at any site on the lattice. For this plot, $\omega_a=\omega_d=0.001$.}
	\label{fig:cargoleave-density-anni}
\end{figure}

In figure  (\ref{fig:cargoleave-density-anni}), we have shown how the total probability 
of cargo detachment at any point on the lattice is influenced by the kinesin density, $r_m$. 
The figure shows that at 
 a low  value of the kinesin-association rate by the cargo, $r_{\rm an}$, 
 the extent of crowding influences  the cargo-association  time to the microtubule only mildly. 
 The situation changes significantly when   the kinesin-association rate  is high. In this 
case, the association time of the cargo to the microtubule, in general,  increases  significantly. 
 Further, for large $r_{\rm an}$, 
   the crowding density of  free  kinesins  affects     the  association  time of the  cargo 
 significantly with the association time being larger for lower crowding density.   

\begin{figure}[!]
	\includegraphics[width=0.5\textwidth]{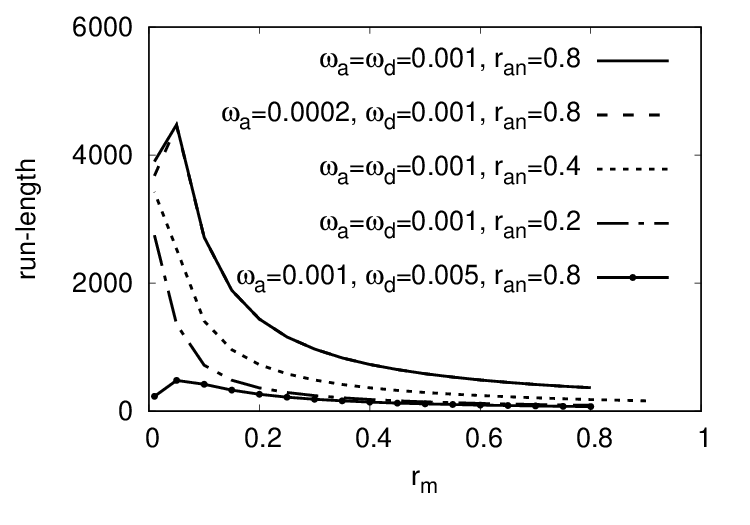}
	\caption{Run-length of the cargo plotted with  crowding density, $r_m$.  
	The cargo can bind at the most two kinesins ($m=2$).  }
	\label{fig:runlength-2kin}
\end{figure}

The dependence of the   run-length of the cargo on the crowding density  
 can be obtained upon  solving equations (\ref{p2})-(\ref{q1}) numerically.  
 Figure (\ref{fig:runlength-2kin}) shows run-length plots for   different  values of  
 the kinesin-association rate, $r_{\rm an}$,  
 and kinesin attachment and detachment rates, $\omega_a$ and $\omega_d$, respectively. 
 For small  $r_{\rm an}$,  the run-length decreases monotonically.
 However, for large $r_{\rm an}$, the run-length increases initially for low crowding. In this case, 
 due to  large $r_{\rm an}$, the cargo benefits from 
 the kinesin-association  process at low crowding. As the 
 crowding density increases, due to limited number of 
  binding sites, the cargo no longer benefits from kinesin association
   and the run-length decreases.  This variation of  the 
   run-length with the crowding density is 
   consistent with earlier experimental predictions \cite{conway}. 
   With the increase in the kinesin detachment rate, $\omega_d$, 
  the run-length  of the cargo  decreases significantly. However, as found earlier, 
 a decrease in the rate of kinesin attachment, $\omega_a$,
  to the cargo has mild effect on   cargo's run-length. 

\begin{figure}[!]
	\includegraphics[width=0.5\textwidth]{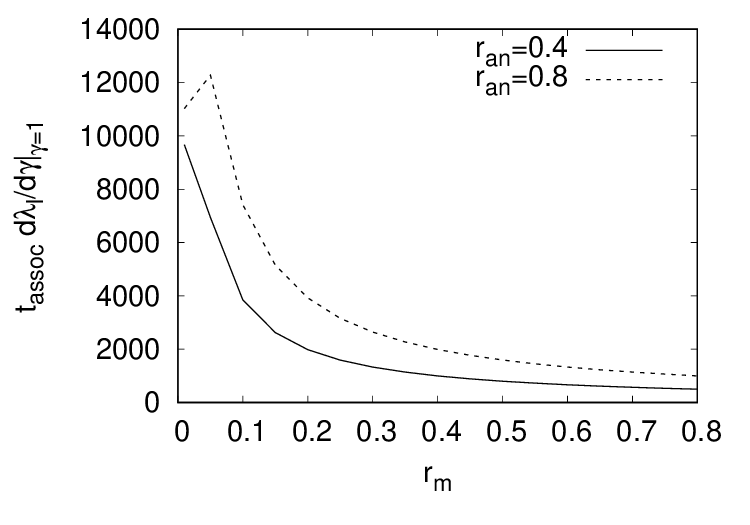}
	\caption{ $y$-axis represents the product of two factors that 
	dominate the nature of the 
	run-length. $t_{\rm assoc}$ is the association time of the cargo and 
	$\frac{d\lambda_l}{d\gamma}$ arises while computing the average 
	velocity of the cargo (see the text). 
	This plot  is for $m=2$ with $\omega_a=\omega_d=0.001$. }
	\label{fig:largest-eigenvalue}
\end{figure}

The fact that the  run-length of the cargo for high kinesin-association rates increases with 
the crowding density initially is consistent with the estimates obtained from the analysis of the largest 
  eigenvalue of the transition matrix $S$ and the association time.  In the limit of large time, the 
  solutions for the probabilities are given by 
  \begin{eqnarray}
  {\bf H}\approx c_1 e^{\lambda_l t} {\bf X},
  \end{eqnarray}
  where $c_1$ is a constant, $\lambda_l$ is the largest of the four eigenvalues 
   of the transition matrix ${\bf S}$ with all of them being negative
  and ${\bf X}=(x_1,\ x_2,\ x_3,\ x_4)^T$ is the corresponding eigenvector. 
   The average distance travelled by the cargo and its average velocity  can be obtained from 
  $
  \langle i\rangle=\sum_{i=-\infty}^{\infty} i [P_1(i,t)+P_2(i,t)+Q_1(i,t)+Q_2(i,t)]\nonumber \\ =\gamma\frac{d}{d\gamma}[\tilde P_1(\gamma,t)+\tilde P_2(\gamma,t)+\tilde Q_1(\gamma,t)+\tilde Q_2(\gamma,t)]\mid_{\gamma=1}
 $ and $v={\langle i\rangle}/{t}$, respectively. 
 In the large time limit, the dominant contribution to the velocity is of the 
 form $v\approx [\gamma c_1 e^{\lambda_l t}\frac{d\lambda_l}{d\gamma} 
  \sum_{i=1}^{4}x_i]\mid_{\gamma=1}$. 
 Using $1/\lambda_l$ as an estimate of the association time, $t_{\rm assoc}$, 
  and finding $\frac{d\lambda_l}{d\gamma}\mid_{\gamma=1}$ 
  numerically for given parameter values, we have plotted 
 $t_{\rm assoc}\frac{d\lambda_l}{d\gamma}\mid_{\gamma=1}$ as a function of the crowding density, $r_m$, 
 in figure (\ref{fig:largest-eigenvalue}). Plots in figure 
 (\ref{fig:largest-eigenvalue}) display similar trends as found in figure  (\ref{fig:runlength-2kin}) for the 
 run-length. 
Although $t_{\rm assoc} v$ gives an estimate of the run-length, the 
 variation in the run-length with the crowding density as seen in figure (\ref{fig:runlength-2kin}) 
 essentially arises from $t_{\rm assoc}\frac{d\lambda_l}{d\gamma}\mid_{\gamma=1}$.  It can be 
 verified numerically that the variation in the remaining factors in $v$ is almost negligible over 
 the entire range of $r_m$, $[0:1]$.  

The dynamical equations for a cargo that can bind at the most three kinesins, i.e.  
$m=3$, are shown  in
Appendix \ref{appendix:model2}. The variation of  $t_{\rm assoc}\frac{d\lambda_l}{d\gamma}\mid_{\gamma=1}$
 with the crowding 
density as obtained from the analysis of the largest eigenvalue is  shown   in figure (\ref{fig:largest-eigenvalue-3kin}). 
  
\begin{figure}[!]
	\includegraphics[width=0.5\textwidth]{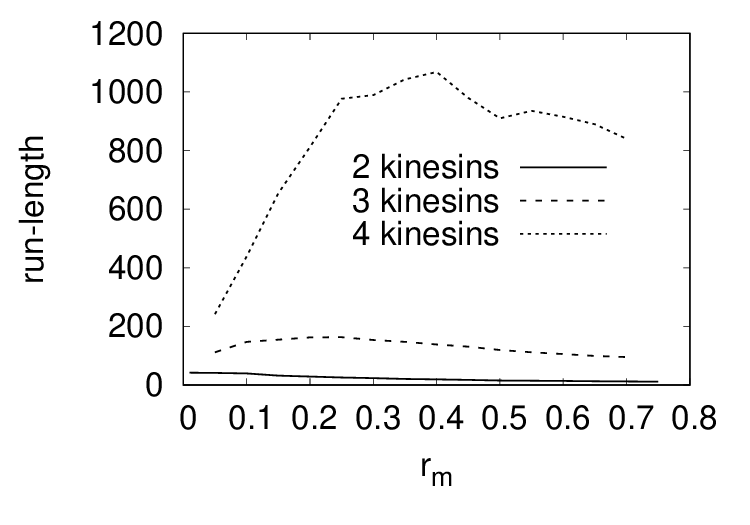}
	\caption{Run-length of the cargo as a function of  free-kinesin density 
	along the microtubule track.
	 For this figure, $\omega_a=\omega_d=0.05$, $r_{an}=0.4$.  The results are obtained upon 
	averaging over $500$ samples for $m=2$ and $3$, and over $1500$ samples for $m=4$. }
	\label{fig:234-runlength}
\end{figure}

Next we simulate the cargo dynamics with the cargo having  $m=2,\ 3, {\rm and}\ 4$ kinesin 
binding sites. 
Figure (\ref{fig:234-runlength}) shows the change 
in  the run-length of the cargo  with  
free-kinesin density, $r_m$,  for  $m=2,\ 3,\ {\rm and}\ 4$. 
 Simulations show an 
  initial increase in the run-length with the free-kinesin-density for $m=3\ {\rm and}\  4$;  
  a trend that was shown earlier in figure (\ref{fig:runlength-2kin}).  

\subsection{Processive motion of free kinesins for $m=3$}
\label{subsec:processive}
In this section, we study the motion of the cargo in the presence of free kinesins which move processively 
on the microtubule track.    In addition, kinesins from 
the intracellular environment can attach to the microtubule and those  walking on the microtubule 
can leave the microtubule at given rates. 

Here we simulate this system using  the cellular automaton method. As before, the 
microtubule is represented by a one-dimensional lattice.
We begin with the cargo positioned at one end of the lattice. The lattice sites 
are randomly occupied by free kinesins with an average density, $r_m$. 
The cargo moves following the 
association  mechanism mentioned  earlier.  
 We assume that the kinesins move unidirectionally in  the same 
direction as that of the cargo. 
The motion of the free kinesins follows the  rules of the paradigmatic totally asymmetric 
simple exclusion process \cite{tasep1}.  Accordingly, each kinesin can 
walk  to the neighbouring site forward  provided  
the target site  is not occupied by another kinesin.  
The attachment and detachment of kinesins are as per the Langmuir kinetics 
considered in \cite{frey-lang}. 
A kinesin can attach to a lattice site at rate $\omega_{a, {\rm kin}}$  provided 
the site is empty and a  free kinesin   can detach from the lattice
at rate $\omega_{d, {\rm  kin}}$. 
A kinesin located at the boundary site can exit from the lattice at   rate, $\beta$. 
We follow  random sequential updating scheme with  probability $p$ for cargo update and $1-p$ 
for updating the rest of the sites. 
Depending on  the site  chosen, the state of the site (or of the cargo) is  updated 
following the aforementioned rules.  The description of various parameters is  provided in  
Appendix \ref{appendix:moving}. 

In figure (\ref{fig:moving}), we have plotted run-lengths for three different scenarios, 
(i)  free kinesins are stalled (static obstacles) (ii) free kinesins are in motion and (iii) free kinesins  
are in motion and they can attach to (or detach from) the microtubule at rates  
$\omega_{a, {\rm kin}}$ (or  $\omega_{d, {\rm  kin}}$).  
Plots  indicate that  in  case of processive free kinesins (case (ii)), the run-length of the cargo 
peaks at a higher crowding density with a higher maximum value as 
compared to the stalled case.   The run-length reduces significantly in case of random 
attachment and detachment of free kinesins (case (iii)). 
Figure (\ref{fig:detail-moving}) in Appendix \ref{appendix:moving} shows that  
the attachment processes  lower  the run-length  significantly. 
 Although,  due to increased   effective crowding density, the cargo 
 remains bound to the  microtubule for a  large span of  time by associating  free kinesins, 
 the crowding restricts the run-length.
As a result of this, the run-length attains its maximum value at a lower value of the 
 crowding density, $r_m$. 
 \begin{figure}[!]
	\includegraphics[width=0.5\textwidth]{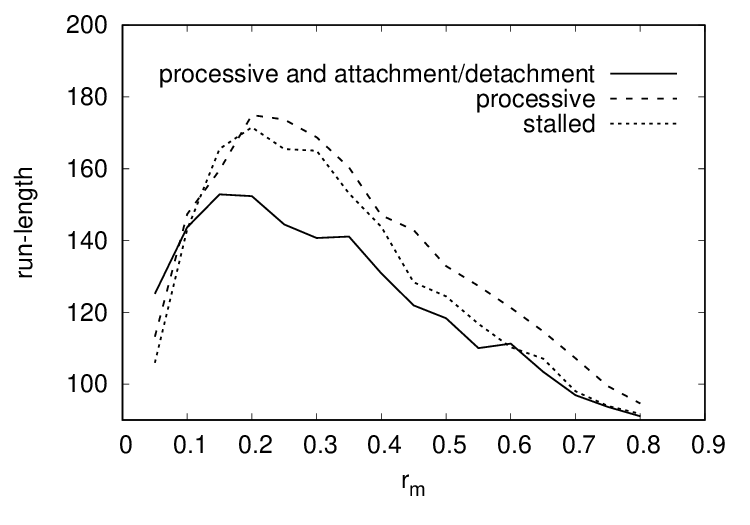}
	\caption{ Run-length of the cargo as a function of  crowding 
	density, $r_m$, under different conditions. For 
	"processive and attachment/detachment", free kinesins move processively along the microtubule 
	and can also randomly attach/detach to/from the microtubule. For "processive" case, free 
	kinesins only move processively without any attachment/detachment dynamics. In the "stalled" case, 
	free kinesins  do not move and  do not  attach or detach from/to the microtubule.
	 For  "processive and attachment/detachment"  plot, 
	$\omega_{a,{\rm kin}}=\omega_{d,{\rm kin}}=0.01$. For the rest of the  cases, 
	$\omega_{a,{\rm kin}}=\omega_{d,{\rm kin}}=0$. Other parameter values are $\omega_a=\omega_d=0.05$, 
	$r_{\rm an}=0.4$, $p=0.1$, and  
	$\beta=0.6$. The total number of lattice sites is $2000$. 
	Run-lengths are obtained upon averaging over $4000$ samples.}
	\label{fig:moving}
\end{figure} 

\begin{figure}[!]
	\includegraphics[width=0.5\textwidth]{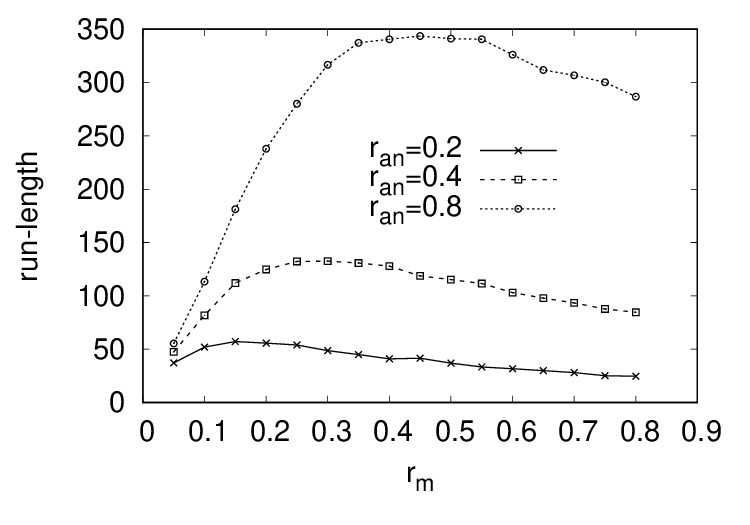}
	\caption{ Run-length of the cargo as a function of   crowding 
	density, $r_m$, for different values of the kinesin-association rate $r_{\rm an}$. Free kinesins move processively and they can randomly attach or detach to or from the microtubule. 
	For this plot $\omega_{a,{\rm kin}}=0.0008,\  \omega_{d,{\rm kin}}=0.0016$ \cite{leduc},  $p=0.1$, and  
	$\beta=0.6$. 
	The total number of lattice sites is $2000$. 
	Run-lengths are obtained upon averaging over $4000$ samples.}
	\label{fig:moving-expt}
\end{figure}

\begin{figure}[!]
	\includegraphics[width=0.5\textwidth]{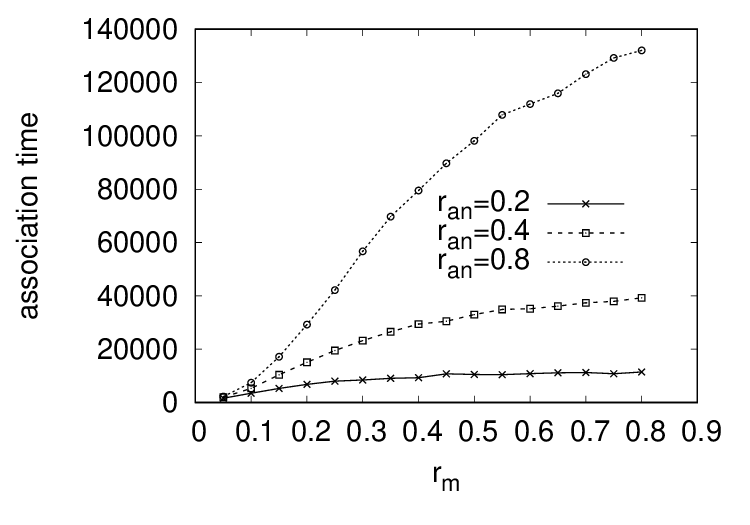}
	\caption{ Association time  of the cargo to the microtubule
	 as a function of  crowding density, $r_m$, 
	for different values of the  kinesin-association rate  ($r_{an}$).   
		Free kinesins move processively and they can randomly attach or detach to or from the microtubule. 
	For this plot $\omega_{a,{\rm kin}}=0.0008,\  \omega_{d,{\rm kin}}=0.0016$,  $p=0.1$, and  
	$\beta=0.6$. 
	The total number of lattice sites is $2000$. 
	 The association time is expressed as the total number of discrete time steps of simulation 
	 till the cargo leaves the microtubule. Association times
	  are obtained upon averaging over $4000$ samples.}
	\label{fig:runtime-expt}
\end{figure}

\begin{figure}[!]
	\includegraphics[width=0.5\textwidth]{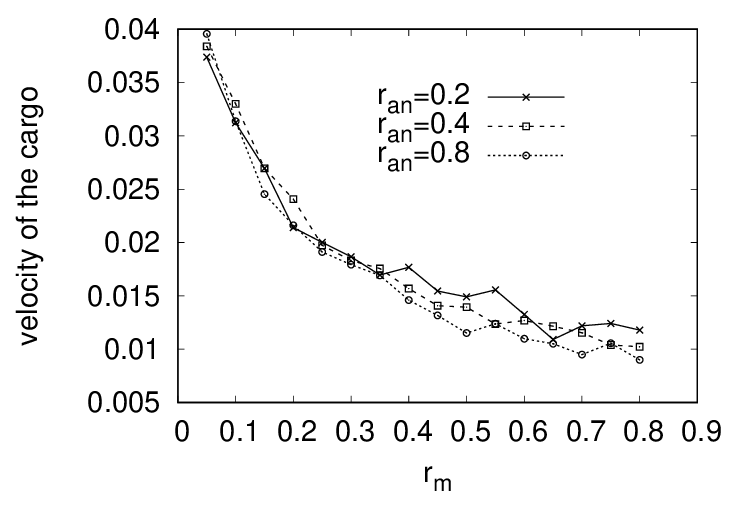}
	\caption{ Velocity of the cargo 
	 as a function of   crowding density, $r_m$, 
	for different values of the  kinesin-association rate  ($r_{an}$).   
		Free kinesins move processively and they can randomly attach or detach to or from the microtubule. 
	For this plot $\omega_{a,{\rm kin}}=0.0008,\  \omega_{d,{\rm kin}}=0.0016$,  $p=0.1$, and  
	$\beta=0.6$. 
	The total number of lattice sites is $2000$. 
	 For a given $r_m$, we have computed  the velocity 
	for every sample.  The values presented  here are obtained upon averaging of $4000$ samples. }
	\label{fig:velocity-expt}
\end{figure}

Figure (\ref{fig:moving-expt}) shows the variation in the run-length   as the 
association rate $r_{\rm an}$ is changed. The peaks in the run-lengths are similar to 
what have been observed earlier. 
Figure (\ref{fig:runtime-expt}) shows a comparison of how the association time 
of the cargo to the microtubule depends on $r_m$ for different 
values of  the kinesin-association rate, $r_{\rm an}$. The association time is 
expressed as   the total number of discrete time steps of simulation till the cargo leaves the 
microtubule. An increase in  $r_{\rm an}$ helps cargo stay attached to the microtubule for a 
longer span of time while as per figure (\ref{fig:velocity-expt}),   the velocity of the cargo 
decreases monotonically with the crowding density, $r_m$. Further, no significant variation in 
 the velocity is seen with  $r_{\rm an}$. 
 The association time and the velocity vary  with $r_m$  in such a manner that their  
 product exhibits a peak at a specific  value of $r_m$.


\pagebreak

\section{Summary}
The motion of cargos on   biopolymeric tracks  crowded due to free or 
cargo-bound motor proteins  is  a subject of immense experimental and theoretical investigations. 
The central goal of these studies is to understand how  cargos manage to 
overcome the motor traffic in order to transport necessary materials in a robust manner.   
Motivated by some of the experimental observations on 
translocation of quantum dot cargos 
in crowded environments, we have modelled mathematically  and computationally 
the motion of a cargo that can  bind kinesins  present along its trajectory 
on the microtubule. 

In the  mathematical modelling, the kinesins on the microtubule track are assumed to be stalled. 
 Besides taking into account the kinesin-association property of the cargo,  
our model incorporates the following dynamical rules.
(i)  The cargo has a limited number of kinesin-binding sites as 
a result of which it can bind at the most a given number of kinesins, (ii) bound kinesins can 
detach from the cargo and kinesins from the 
intracellular space can bind to the cargo at certain rates and  (iii) the cargo leaves the 
microtubule if all the kinesins detach from the cargo.
Upon finding the cargo velocity for a   toy model where the cargo never leaves the 
microtubule and keeps moving  forward by removing obstacles via association, we 
generalize the mathematical formulation  to take into account the aforementioned 
aspects of the cargo dynamics.  
We show that the two features, namely,  the  
crowding along the microtubule and the  ability of the cargo to associate kinesins 
have competing effects on the run-length of the cargo. For low crowding density, 
as  the crowding density 
increases, the cargo benefits  due to its ability to associate kinesins. This leads to 
an increase in the  run-length with the crowding density. However,  as the 
crowding density increases further, due to 
its limited number of kinesin binding sites, the cargo does not benefit anymore 
through kinesin association. As a consequence,   the run-length decreases for large values  
of  the crowding density.  This nature of the run-length has been predicted  earlier from 
experimental observations. We show that this property of the run-length is governed 
 by the largest eigenvalue of the transition matrix describing the 
 dynamics of the cargo. The  model can be generalized further   to incorporate   other 
 features such as reversals of the cargo, bidirectional movements of the cargo, 
 pausing of the cargo  etc.  with the frequencies of such  events  depending on
 the crowding  density.  
 The  present  work  lays  a foundation for such studies. Additionally, this analysis may 
 also lead to testable predictions for cargo's motile properties 
  once  appropriate   parameter values are available.   
  
Next, we have simulated cargo transport with 
 processive motion of free kinesins as well as  binding and unbinding 
   of motors to or from the 
 microtubule. For different values of the rate of kinesin association to the  cargo, 
 the  run-lengths show prominent peaks as the crowding density is changed. 
 However, overall, the run-length  decreases significantly due to  binding of motors 
 to the microtubule, a process that increases the effective crowding density. 
 As a consequence of cargo's ability to associate kinesin, the  
 association-time of the cargo to the microtubule increases with the increase in the crowding density. 
 The velocity of the cargo, on the other hand,  decreases with the crowding density and it remains 
 approximately unchanged with the kinesin-association rate of the cargo. 
 Incorporating the processive motion in the mathematical model would add another 
 level of complexity which can be a subject of future studies.

\appendix
\section{Model 1} \label{appendix:model1}
\begin{figure}[!]
	\includegraphics[width=0.45\textwidth]{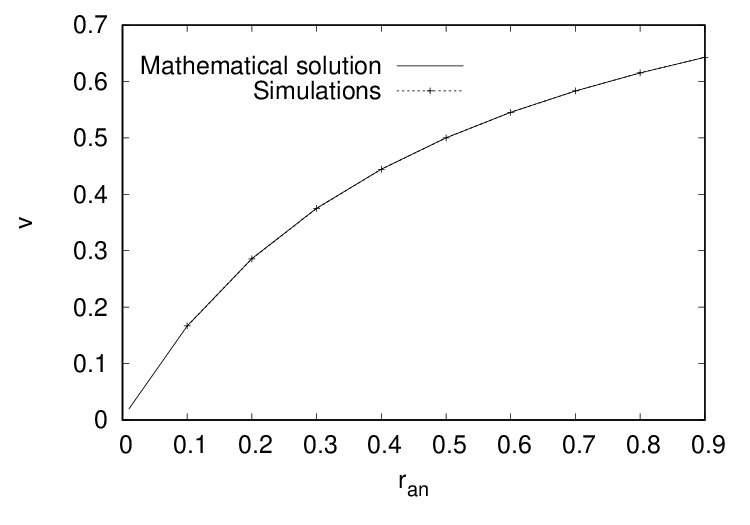}
	\caption{Velocity of the cargo as a function of $r_{an}$ with  $r_m=1/2$. }
	\label{fig:single}
\end{figure}

In  the matrix form the differential equations  (\ref{Peq}) and (\ref{Qeq}) 
 appear as 
 \begin{eqnarray}
 \frac{d}{dt}{\bf G}(\gamma,t)={\bf R} {\bf G}(\gamma,t),
 \end{eqnarray}
 where 
 \begin{eqnarray}
 {\bf G}(\gamma,t)=
  {\begin{pmatrix}
   \tilde P(\gamma,t)\\
   \tilde Q(\gamma,t)\\
  \end{pmatrix} }  \ {\rm and }\ \nonumber \\
  {\bf R}={\begin{pmatrix}
   & (1-r_m) \gamma-1 & r_{an}\\
   &r_m \gamma & -r_{an}\\
  \end{pmatrix}. }
  \end{eqnarray}
Here ${\bf R}$ is a transition matrix. For $\gamma=1$, the sum of all the elements in a column is $0$.
One may find out the solutions of these equations upon finding the eigenvalues and eigenvectors 
  of ${\bf R}$. The eigenvectors corresponding to the eigenvalues $\lambda_{\pm}$ are, respectively, 
  \begin{eqnarray}
  {\begin{pmatrix}
   & 1\\
   &\frac{\lambda_++1-(1-r_m)\gamma}{r_{an}}\\
  \end{pmatrix} }\ \  {\rm and}\ \  
  {\begin{pmatrix}
   & 1\\
   &\frac{\lambda_-+1-(1-r_m)\gamma}{r_{an}}\\
  \end{pmatrix} },
  \end{eqnarray}
  where $\lambda_{+,-}=\frac{1}{2}[-(r_{an}+1-(1-r_m) \gamma)\pm A]$ with 
  $A=\sqrt{(r_{an}+1-(1-r_m) \gamma)^2-
  4 r_{an} (1-\gamma)}$.
  The solutions for the generating functions are 
  \begin{eqnarray}
  {\begin{pmatrix}
   & \tilde P(\gamma,t)\\\
   &\tilde Q(\gamma,t)\\
  \end{pmatrix} }=c_1e^{\lambda_+t}{\begin{pmatrix}
   & 1\\
   &\frac{\lambda_++1-(1-r_m)\gamma}{r_{an}}\\
  \end{pmatrix} }+\nonumber\\ 
  c_2e^{\lambda_-t}{\begin{pmatrix}
   & 1\\
   &\frac{\lambda_-+1-(1-r_m)\gamma}{r_{an}}\\\end{pmatrix} },
\end{eqnarray}
where  $c_1,  \ c_2$ are integration constants. 
We consider the initial conditions $P(i,t=0)=Q(i,t=0)=1/2$ for $i=0$. 
Using these conditions, we find  
\begin{eqnarray}
c_1=\frac{1}{2A}\left[r_{an}-\lambda_- - 1+(1-r_m)\gamma\right] \ \ \ {\rm and} \nonumber\\ 
c_2=\frac{1}{2}-c_1=\frac{1}{2A}\left[A-r_{an}+\lambda_-+1-(1-r_m)\gamma\right].
\end{eqnarray}
Since in  the large time limit, the solutions are governed by the largest eigenvalue, we have   
\begin{eqnarray}
\tilde P(\gamma,t)+\tilde Q(\gamma,t)\approx c_1 
e^{\lambda_+t}\left[1+\frac{\lambda_++1-(1-r_m)\gamma}{r_{an}}\right].\nonumber\\
\end{eqnarray}
Upon taking derivatives of $\tilde P(\gamma,t)+\tilde Q(\gamma,t)$ with respect to $\gamma$, we have 
\begin{eqnarray}
&&\langle i\rangle=\left[\gamma \left(\frac{d\tilde P}{d\gamma}+\frac{d\tilde Q}{d\gamma}\right)\right]_{\gamma=1}\nonumber\\
&&=\bigg\{\gamma \frac{dc_1}{d\gamma}e^{\lambda_+t}\big(1+\frac{\lambda_++1-(1-r_m)\gamma}{r_{an}}\big)\bigg\}_{\gamma=1}+\nonumber\\
&&\bigg\{\gamma c_1 e^{\lambda_+t}\frac{d\lambda_+}{d\gamma} t\big(1+\frac{\lambda_++1-(1-r_m) \gamma}{r_{an}}\big)\bigg\}_{\gamma=1}+\nonumber\\
&&\bigg\{\gamma c_1 e^{\lambda_+t}\frac{1}{r_{an}}\big(\frac{d\lambda_+}{d\gamma}-(1-r_m)\big)\bigg\}_{\gamma=1}.
\end{eqnarray}
In the large time limit,  we finally have 
 \begin{eqnarray}
 \langle i\rangle/t=\frac{d\lambda_+}{d\gamma}\mid_{\gamma=1}.
 \end{eqnarray}
Using
 \begin{eqnarray}
 \frac{d\lambda_+}{d\gamma}\bigg |_{\gamma=1}=\left[\frac{1-r_m}{2}+\frac{1}{2}\frac{dA}{d\gamma}\right]_{\gamma=1},
 \end{eqnarray}
 where 
 \begin{eqnarray}
 \frac{dA}{d\gamma}\bigg |_{\gamma=1}=\frac{r_{\rm an}+r_{\rm an}r_m-r_m+r_m^2}{r_{\rm an}+r_m},
 \end{eqnarray}
 we have 
\begin{eqnarray}
 v=\frac{d\lambda_+}{d\gamma}\mid_{\gamma=1}=\frac{r_{an}}{r_{an}+r_m}. \label{v-one-kinesin}
 \end{eqnarray}
 Figure (\ref{fig:single}) shows plots of  velocity obtained mathematically and through simulations. 
 
 \section{Model 2}  \label{appendix:model2}
 \subsection{$m=3$}
\begin{figure}[ht!]
	\includegraphics[width=0.5\textwidth]{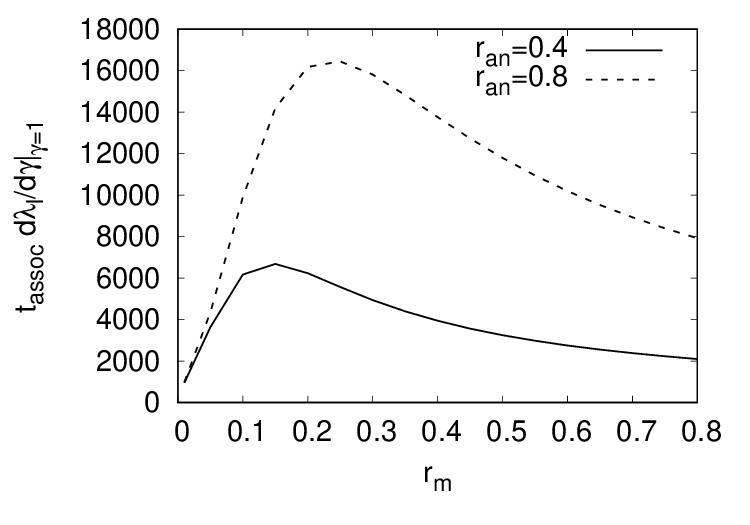}
	\caption{ $y$-axis represents the product of two factors that 
	dominate the nature of the 
	run-length. $t_{\rm assoc}$ is the association time of the cargo and 
	$\frac{d\lambda_l}{d\gamma}$ arises while computing the average 
	velocity of the cargo (see the text). This plot is 
	for $m=3$ with $\omega_a=\omega_d=0.01$. }
	\label{fig:largest-eigenvalue-3kin}
\end{figure}

 A cargo  that can bind at the most three kinesins  can be in four possible states, namely, 
 bound to one, two or three kinesins or not bound to  any kinesin. 
 Possible configurations of two neighbouring sites can be of 
 $P$ type or $Q$ type depending on whether the site in front of the cargo is occupied by a free kinesin or 
 empty. For example,  $P_n(i,t)$ ($n=1, \ 2,\ {\rm or}\  3$) indicates the probability of 
 a configuration where  
 a cargo, bound to $n$ number of kinesins,  
 is present at the $i$-th site at time $t$ while the $(i+1)$-th site is
 empty. Similarly,  $Q_n(i,t)$ ($n=1, \ 2,\  {\rm or}\ 3$) represents the probability of 
 a  configuration where 
 a cargo, bound to $n$ number of kinesins,  is present at the $i$-th site at time $t$ while the $(i+1)$-th site is
 occupied by a free kinesin. The change in these probabilities with time is described by the equations
 \begin{widetext}
\begin{eqnarray}
&&\frac{d}{dt}P_3(i)=(1-r_m) P_3(i-1)-P_3(i)+r_{an}Q_2(i)-\omega_d P_3(i)+\
\omega_aP_2(i),\label{p33}\\
&&\frac{d}{dt}P_2(i)=(1-r_m) P_2(i-1)-P_2(i)+r_{an}Q_1(i)+\omega_d P_3(i)+
\omega_aP_1(i)-(\omega_a+\omega_d) P_2(i),\label{p23}\\
&&\frac{d}{dt}P_1(i)=(1-r_m) P_1(i-1)-P_1(i)+\omega_d P_2(i)-
(\omega_a+\omega_d) P_1(i),\label{p13}\\
&&\frac{d}{dt}Q_3(i)=r_m P_3(i-1)+\omega_a Q_2(i)-\omega_d Q_3(i),\label{q33}\\
&&\frac{d}{dt}Q_2(i)=r_m P_2(i-1)+\omega_a Q_1(i)+
\omega_d Q_3(i)-(\omega_d +\omega_a)Q_2(i)-r_{\rm an} Q_2(i), \ \ {\rm and}\label{q23}\\
&&\frac{d}{dt}Q_1(i)=r_m P_1(i-1)-r_{an} Q_1(i)+\omega_d Q_2(i)-(\omega_a+\omega_d)Q_1(i).\label{q13}
\end{eqnarray}
\end{widetext}
Additionally, as in $m=2$ case, we have
\begin{eqnarray}
\frac{d}{dt} P_0(i)=\omega_d P_1(i)\label{p0}\  {\rm and}\  
\frac{d}{dt} Q_0(i)=\omega_d Q_1(i) \label{q0}.
\end{eqnarray} 
\begin{widetext}
Defining the generating functions as $\tilde P_n(\gamma,t)=\sum_{i=-\infty}^{\infty} \gamma^i P_n(i,t)$  and 
$\tilde Q_n(\gamma,t)=\sum_{i=-\infty}^{\infty} \gamma^i Q_n(i,t)$, we have 
\begin{eqnarray}
\frac{d}{dt} {\bf H}(\gamma,t)={\bf S} {\bf H}(\gamma,t), 
\end{eqnarray}
where ${\bf H}$ is a column matrix 
\begin{eqnarray}
{\bf H}(\gamma,t)={\begin{pmatrix}
& \tilde P_3(\gamma,t)\\
 & \tilde P_2(\gamma,t)\\
 &\tilde P_1(\gamma,t)\\
 &\tilde Q_3(\gamma,t)\\
 & \tilde Q_2(\gamma,t)\\
 &\tilde Q_1(\gamma,t)\\
 \end{pmatrix} },
  \end{eqnarray}
  and ${\bf S}$ is a $6\times 6$ matrix 
  \begin{eqnarray}
  {\bf S}={\begin{pmatrix}
   & (1-r_m) \gamma-\omega_d -1& \omega_a \  &0 \ & 0\   & r_{an} & 0 \\
 &\omega_d & (1-r_m) \gamma-\omega_a - \omega_d-1 & \omega_a&  0 & 0 & r_{\rm an}\\
 &0 &\omega_d & (1-r_m)\gamma-\omega_a-\omega_d-1 &0 &0 & 0\\
  & r_m\gamma & 0  & 0 & -\omega_d & \omega_a & 0 \\
 &0  & r_m\gamma  & 0 & \omega_d   & -\Omega & \omega_a \\
 &0 &0  & r_m\gamma & 0 & \omega_d &-\Omega\\ 
  \end{pmatrix} },
 \end{eqnarray}
 where $\Omega=(\omega_a+\omega_d+r_{\rm an}).$
 \end{widetext}
 As in case of $m=2$, here again  the variation in the run-length is governed by the quantity 
 $t_{\rm assoc}\frac{d\lambda_l}{d\gamma}\mid_{\gamma=1}$ where $\lambda_l$ is the largest 
 eigenvalue of matrix $S$. Figure (\ref{fig:largest-eigenvalue-3kin}) shows the variation in the  
 $t_{\rm assoc}\frac{d\lambda_l}{d\gamma}\mid_{\gamma=1}$ with $r_m$.
 

 \section{Processive movement of free kinesins}  \label{appendix:moving}
 Descriptions of parameters  used  in  simulations 
 are provided in  table \ref{table:parameter}.  
 \begin{table}[ht!]
\label{table:parameter}
\begin{tabular}{|c|c|c|}
\hline
 $\Delta t$ &$d/v [s]$ &\makecell{discrete time step, \\ $v$  [nm $s^{-1}$] - velocity of free kinesin/cargo,\\  
 $d$   [nm] - length \\ of the tubulin dimer (lattice spacing)} \\
 \hline
 $\omega_{a,kin}$ &$\bar{\omega}_{a,kin}d \Delta t$& \makecell{dimensionless\\  kinesin attachment rate}\\
  \hline
 $\omega_{d,kin}$ &  $\bar{\omega}_{d,kin}\Delta t$&  \makecell{dimensionless\\ kinesin detachment rate}\\
 \hline
 $\omega_a$ & $-$ & \makecell{dimensionless \\ kinesin attachment rate to cargo}\\
 \hline
 $\omega_d$ & $-$ & \makecell{dimensionless kinesin\\ detachment rate from cargo}\\
 $r_{\rm an}$ & $-$ & \makecell{dimensionless  association rate\\ of free kinesin  to cargo}\\
 \hline
 \end{tabular}
 \caption{Description of parameters used in simulations in section \ref{subsec:processive}. 
 A tubulin dimer is the basic subunit of a microtubule.
 $\bar{\omega}_{a,kin}$ and $\bar{\omega}_{d,kin}$.  
 $\bar{\omega}_{a,kin}$ denotes the attachement rate of a free kinesin to the microtubule 
 per unit length per  unit time [$\mu m^{-1}\ s^{-1}$] and 
  $\bar{\omega}_{d,kin}$ denotes the detachment rate  of a free kinesin from the  microtubule 
  per unit time [$s^{-1}$]. }
\label{table:parameter}
\end{table}

\begin{figure}[ht!]
	\includegraphics[width=0.5\textwidth]{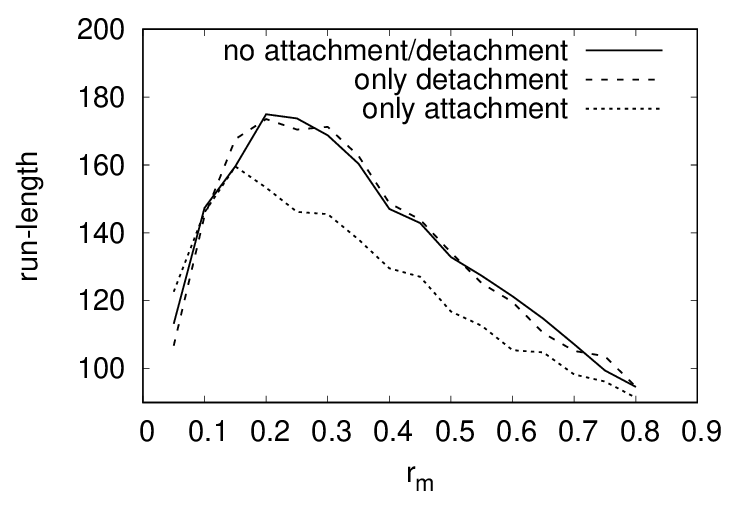}
	\caption{ $\omega_a=\omega_d=0.05$, 
	$r_{\rm an}=0.4$, $p=0.1$, and  
	$\beta=0.6$. The total number of lattice sites is $2000$. Run-lengths are obtained upon 
	averaging over $4000$ samples.  In case  of only attachment, $\omega_{a,{\rm kin}}=0.01$.
	 In case of only detachment, $\omega_{d,{\rm kin}}=0.1$.}
	\label{fig:detail-moving}
\end{figure}

 Figure (\ref{fig:detail-moving}) shows how the run-length varies with the crowding density  
 in the  three cases -   Processive movement  of free kinesins and 
(i)    binding of kinesins   to the microtubule at rate  $\omega_{a, {\rm kin}}$,
(ii)   unbinding of kinesins  from the microtubule at rate $\omega_{d, {\rm kin}}$, and (iii) no 
binding  or unbinding of kinesins to or from the microtubule.  

\clearpage
 
\end{document}